\documentclass[conference]{IEEEtran}

\usepackage{algorithm,algorithmic}
\usepackage{amsmath,amssymb,mathrsfs}
\usepackage{cite}
\usepackage{subfigure}
\usepackage{graphicx,color}

\usepackage{hyperref}
\usepackage{subfigure}
\usepackage{balance}
\usepackage{tikz,siunitx}
\usepackage{url}

\newtheorem{thm}{Theorem}
\newtheorem{cor}{Corollary}[thm]

\definecolor{sblue}{RGB}{0,51,120}
\definecolor{sred}{RGB}{200,51,130}

\newcommand{\matc}[1]{\mbox{\boldmath $\mathcal{#1}$}}

\ifCLASSINFOpdf
\else
\fi

\begin{document}

\title{Network-ELAA Beamforming and Coverage Analysis for eMBB/URLLC in Spatially Non-Stationary Rician Channels}

\author{Jinfei Wang$^{1}$, Yi Ma$^{1\dagger}$, Na Yi$^{1}$, Rahim Tafazolli$^{1}$, and Fan Wang$^{2}$\\
	{\small $^{1}$5GIC and 6GIC, Institute for Communication Systems, University of Surrey, Guildford, UK, GU2 7XH}\\
		{\small $^{1}$Emails: (jinfei.wang, y.ma, n.yi, r.tafazolli)@surrey.ac.uk}\\
		{\small $^{2}$Huawei UK, Email: (fan.wang)@huawei.com}
		}
\markboth{}%
{}

\maketitle

\begin{abstract}
In vehicle-to-infrastructure (V2I) networks, a cluster of multi-antenna access points (APs) can collaboratively conduct transmitter beamforming to provide data services (e.g., eMBB or URLLC).  
The collaboration between APs effectively forms a networked linear antenna-array with extra-large aperture (i.e., network-ELAA), where the wireless channel exhibits spatial non-stationarity. 
Major contribution of this work lies in the analysis of beamforming gain and radio coverage for network-ELAA non-stationary Rician channels considering the AP clustering. 
Assuming that: {1)} the total transmit-power is fixed and evenly distributed over APs, {2)} the beam is formed only based on the line-of-sight (LoS) path, it is found that the beamforming gain is concave to the cluster size. 
The optimum size of the AP cluster varies with respect to the user's location, channel uncertainty as well as data services.
A user located farther from the ELAA requires a larger cluster size. 
URLLC is more sensitive to the channel uncertainty when comparing to eMBB, thus requiring a larger cluster size to mitigate the channel fading effect and extend the coverage. Finally, it is shown that the network-ELAA can offer significant coverage extension ($50\%$ or more in most of cases) when comparing with the single-AP scenario. 
\end{abstract}

\section{Introduction}\label{secI}
Network-ELAA is a special paradigm of extra-large aperture array (ELAA), where a set of multi-antenna access points (APs) are networked either through cable or wireless links and collaboratively provide data services to their mobile users \cite{BJORNSON20193}. Different from networked multi-antenna systems (i.e., network-MIMO) in the conventional sense \cite{4487170}, network-ELAA users are located in the near field, and thus their wireless channels are spatially non-stationary. With the evolution of 5G or beyond 5G framework, network-ELAA has recently gained increasing interests and applications. One of important use-cases is in the vehicle-to-infrastructure (V2I) scenario, where a cluster of roadside units (RSUs) can collaboratively serve their users on the road. When the cluster size is not too large, the RSU cluster can be viewed as a networked non-uniform linear array (non-ULA). Channel measurement campaign results have shown that the V2I wireless channel is Rician distributed, and the Rician K-factor is dependent on the specific wireless environment \cite{6899647}. Due to the channel spatial non-stationarity, RSUs at different physical locations can contribute very differently to the wireless signal transmission. For instance when a RSU cluster is forming a beam onto a downlink user, the beamforming gain might be more sensitive to RSUs located closer to the normal between the user and the cluster and less sensitive to those RSUs at the far end. Certainly, this is for now just a hypothesis, which is yet to be rigorously studied. 
Besides, RSUs are networked mainly through wireless links, and for this reason the exchange of local transmitter-side channel state information (T-CSI) amongst all involved RSUs is not a favorable solution in practice. Then, it would be interesting to investigate a network-ELAA beamforming approach with minimum requirement to the wireless backhaul, and how the beamforming gain varies with respect to the user's location (the coverage problem).
All of these problems motivated this work. 

\begin{figure}[t]
	\centering
	\includegraphics[scale=0.2]{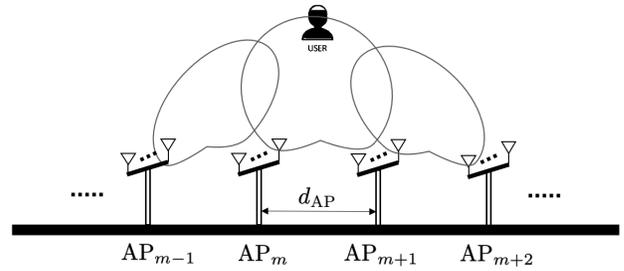}
	\caption{Illustration of the network-ELAA system and collaborative beamforming.}
	\vspace{-2em}
	\label{fig1}
\end{figure}
Our investigation is focused on the theoretical model of the network-ELAA depicted in Fig. \ref{fig1}. 
Basically, the network-ELAA consists of a set of APs, which are placed on a line with equal spacing. Each AP has a number of transmit-antennas forming a small ULA. The network-ELAA wireless channel is spatially non-stationary Rician fading, which is a special case of the statistical non-stationary channel model presented in \cite{liu2021nonstationary}. Provided that APs are connected through capacity-constrained wireless backhaul, this work is interested in the geo-location aware analog beamforming technique, where the beam is formed mainly based on line-of-sight (LoS) paths (e.g., \cite{8580983,alrabeiah2020viwi}), as such avoids the exchange of T-CSI between APs and minimizes the load of wireless backhaul. 
In order to demonstrate the beamforming gain with respect to the cluster size, it is assumed that the total transmit-power is fixed and evenly distributed over APs. 
In summary, our work results in a couple of interesting findings: 

{\em 1)} In the network-ELAA system, the beamforming gain is concave to the size of AP cluster (i.e., the number of APs). In other words, the beamforming gain does not grow monotonically with the number of APs, and there exists an optimum size of the cluster. This observation is very different from that in conventional MIMO/network-MIMO systems where wireless channels are often assumed to be stationary. 

{\em 2)} The optimum size of AP cluster is dependent on the user's location, channel uncertainty and data services. A user located farther from the ELAA requires a larger cluster size to achieve a satisfied beamforming gain. In terms of data services, ultra-reliable low-latency communication (URLLC) requires a larger cluster size than enhanced mobile broadband (eMBB). This is because URLLC is more sensitive to the channel uncertainty and requires more antennas to mitigate the channel fading effect and extend the coverage.

The above findings have been carefully verified through both mathematical analysis and computer simulations. In addition, the two-dimensional (2D) coverage probability is also evaluated for both eMBB and URLLC services using the optimized beamforming gain.

\section{System Model and Problem Statement}\label{secII}
\subsection{Signal Transmission Model}
Consider a network-ELAA system depicted in Fig. \ref{fig1}, where a set of APs ($L$) are placed on a line with the equal spacing ($d_\textsc{ap}$).
Each AP has a set of transmit antennas ($M_\textsc{ap}$) forming a ULA with the antenna spacing of $(\lambda)/(2)$, where $\lambda$ is the wavelength of the carrier ($\lambda\ll d_\textsc{ap}$).
APs are networked in the sense that they can collaboratively form a beam for a narrowband user located on the 2D plane.
Throughout this paper, it is assumed that users share the wireless medium orthogonally either in the time or frequency, thus forming a noise-limited communication model. 

Denote $s$ to be the information-bearing symbol sent by the ELAA with $\mathbb{E}(s)=0$, $\mathbb{E}(ss^*)=1$, where $[\cdot]^*$ stands for the conjugate and $\mathbb{E}(\cdot)$ for the expectation. The transmitted signal is an $(M)\times(1)$ vector, $\mathbf{x}$, specified by ($M\triangleq LM_\textsc{ap}$) 
\begin{equation}\label{eq01}
\mathbf{x}\triangleq\sqrt{P_o}\mathbf{w}s,
\end{equation}
where $\mathbf{w}$ is an $(M)\times(1)$ beamforming vector with the element-wise normalization, and $P_o$ is the transmit power. Assuming that the receiver has a single receive-antenna, the received signal at the baseband in its discrete-time equivalent form reads as
\begin{equation}\label{eq02}
y=\sqrt{P_o}\mathbf{h}^T\mathbf{w}s+v,
\end{equation}
where $\mathbf{h}$ stands for the $(M)\times(1)$ channel vector, $v\sim\mathcal{CN}(0,\sigma^2)$ for the additive white Gaussian noise (AWGN), and $[\cdot]^T$ for the transpose. The instantaneous signal-to-noise ratio (iSNR) is computed by
\begin{equation}\label{eq03}
\mathrm{isnr}=\frac{\beta P_o}{\sigma^2},~\beta=|\mathbf{h}^T\mathbf{w}|^2,
\end{equation}
where $|\cdot|$ stands for the absolute value. 

\subsection{ELAA Channel Model and Location-Aware Beamforming}
The network-ELAA channel is spatially non-stationary and Rician distributed. It is a special case of the LoS and non-LoS mixed non-stationary channel model in \cite{liu2021nonstationary}, which has the following mathematical form
\begin{equation}\label{eq04}
h_m=\alpha_m\left(\sqrt{\frac{\kappa}{1+\kappa}}\phi_m+\sqrt{\frac{1}{1+\kappa}}\nu_m\right),~_{m=0,...,M-1},
\end{equation}
where $h_m$ is the $m$-th element of $\mathbf{h}$, $\kappa$ the Rician K-factor, $\nu_m$ the complex Gaussian noise due to non-LoS paths ($\nu_\mathrm{m}\sim\mathcal{CN}(0,1)$), $\alpha_m$ the path-loss attenuation ($\alpha_m>0$), and $\phi_m=\exp\left(-j\frac{2\pi d_m}{\lambda}\right)$ ($d_m$ is the distance between the $m$-th antenna and the user).

The geo-location aware beamforming requires the knowledge of $\phi_m$ at the network-ELAA, which is only related to the distance $d_m$. 
Once the user's location is known, every AP can individually compute the parameter $\phi_m$, and thus there is no need of exchanging their local T-CSI. Defining a row vector $\matc{\phi}^T\triangleq[\phi_0, ...,\phi_{M-1}]^T$, the beamforming vector is formed by
\begin{equation}\label{eq05}
\mathbf{w}=\matc{\phi}^H,
\end{equation}
where $[\cdot]^H$ stands for the Hermitian transpose.
Applying \eqref{eq04} and \eqref{eq05} into \eqref{eq03} yields
\begin{equation}\label{eq06}
\beta=|a+b|^2,
\end{equation}
where
\begin{equation}\label{eq07}
a=\sqrt{\frac{\kappa}{1+\kappa}}\left(\sum_{m=0}^{M-1}\alpha_m\right),
\end{equation}
\begin{equation}\label{eq08}
b=\sqrt{\frac{1}{1+\kappa}}\left(\sum_{m=0}^{M-1}\alpha_m\nu_m\phi_m^*\right).
\end{equation}

\begin{thm}\label{thm01}
Suppose: c1) $\mathbb{E}(\mathbf{x}^H\mathbf{x})=P$; c2) $\alpha_m=\alpha_0$ (constant). For $M\rightarrow\infty$, the iSNR term in \eqref{eq03} increases asymptotically to
\begin{equation}\label{eq09}
\lim_{M\rightarrow\infty}\mathrm{isnr}=\Big(\frac{\kappa}{1+\kappa}\Big)\left(\frac{P}{\sigma^2}\right).
\end{equation}
\end{thm}
\begin{IEEEproof}
See Appendix \ref{appdx1}.
\end{IEEEproof}
With the condition c2), the iSNR increases with $M$ (or equivalently with the cluster size $L$). 
However, the condition c2) does not hold in the non-stationary Rician channel. 
Therefore, it would be interesting to develop in-depth understanding about the relationship between the iSNR and the cluster size in the non-stationary Rician channel, and if possible determine the optimum cluster size that maximizes the iSNR. 
Moreover, it would be very useful to understand the coverage of eMBB and URLLC with the optimized cluster size. 

\section{Optimization of the Cluster Size}\label{secIII}
\subsection{The Beamforming Gain and Properties}
\eqref{app01} shows that the iSNR is a linear function of the beamforming gain defined by: $\bar{\beta}\triangleq(\beta)/(M)$. 
Denote $\alpha_\bot$ to be the path-loss attenuation corresponding to the shortest distance between the ELAA and the user ($d_{\bot}$). 
For the non-stationary Rician channel, the beamforming gain can be written into the following form
\begin{equation}\label{eq10}
\bar{\beta}=\frac{\alpha_\bot^2}{(1+\kappa)}\Big|\sqrt{\frac{\kappa}{M}}\sum_{m=0}^{M-1}\Big(\frac{\alpha_m}{\alpha_{\bot}}\Big)+\sqrt{\frac{1}{M}}\sum_{m=0}^{M-1}\Big(\frac{\alpha_m}{\alpha_\bot}\Big)\nu_m\phi_m^*\Big|^2.
\end{equation}
Given the condition $d_\textsc{ap}\gg\lambda$, we can assume $\alpha_m$ to be identical for co-located transmit-antennas. 
Then, we have
\begin{equation}\label{eq11}
\frac{1}{\sqrt{M}}\sum_{m=0}^{M-1}\Big(\frac{\alpha_m}{\alpha_\bot}\Big)=\sqrt{\frac{M_\textsc{ap}}{L}}\sum_{l=0}^{L-1}\Big(\frac{\alpha_l}{\alpha_\bot}\Big).
\end{equation}
Denoting $(2q)$ to be the path-loss exponent, we can further have
\begin{equation}\label{eq12}
\Big(\frac{\alpha_l}{\alpha_\bot}\Big)=\Big(\frac{d_\bot}{d_l}\Big)^q=\Big(\frac{d_\bot}{\sqrt{d_\bot^2+l^2d_\textsc{ap}^2}}\Big)^q.
\end{equation}
Applying \eqref{eq12} into \eqref{eq11}, we are interested in the following function
\begin{equation}\label{eq13}
\mathcal{G}(L)\triangleq\frac{\sum_{l=0}^{L-1}(1+l^2\eta)^{-\frac{q}{2}}}{\sqrt{L}}
,~\eta\triangleq\Big(\frac{d_\textsc{ap}}{d_\bot}\Big)^2.
\end{equation}
Generally, we cannot easily claim the convexity or monotonicity of $\mathcal{G}(L)$ as it depends on the parameters $\eta$ and $q$. 
Nevertheless, we find the following result useful to understand the convexity or monotonicity of $\mathcal{G}(L)$.
\begin{thm}\label{thm02}
For the discrete-time function $\mathcal{G}(L)$, it is concave to $L$ when 
\begin{equation}\label{eq14}
\eta<(\sqrt{2}+1)^{\frac{2}{q}}-1;
\end{equation}
or otherwise, $\mathcal{G}(L)$ is a monotonically decreasing function.
\end{thm}
\begin{proof}
See Appendix \ref{appdx2}.
\end{proof}

{\em Theorem} \ref{thm02}, as well as its proof, implies that the convexity of $\mathcal{G}(L)$ depends on the relationship between $d_\textsc{ap}$ and $d_\bot$. When $d_\textsc{ap}$ is fixed, users located closer to the ELAA have a larger $\eta$. 
Then, $\mathcal{G}(L)$ becomes (or is closer to) a monotonically decreasing function, where $\mathcal{G}(L)$ reaches its maximum at a smaller $L$ or even at $L=1$. When users move away from the ELAA, $\eta$ will quickly decrease. This renders $\mathcal{G}(L)$ a concave function. 
In this case, $\mathcal{G}(L)$ is approximately an inverse Gaussian distribution function \cite{10480942}, with its maximum appearing at a larger $L$. 
This phenomenon is also confirmed by our simulation results. 

\begin{cor}\label{cor1}
For a sufficiently large $M$, the beamforming gain $\bar{\beta}$ has the following approximate form, which shows the same convexity or monotonicity as $\mathcal{G}(L)$
\begin{equation}\label{eq15}
\bar{\beta}\approx\frac{\kappa\alpha^2_\bot}{(1+\kappa)M_\textsc{ap}}\mathcal{G}^2(L).
\end{equation}
\end{cor}
\begin{IEEEproof}
The proof of {\em Corollary} \ref{cor1} is rather straightforward, as for a sufficiently large $M$, the random term in \eqref{eq10} is averaged out. Then, \eqref{eq15} can be obtained by applying \eqref{eq13} into the approximated version of \eqref{eq10}.
Since $\mathcal{G}(L)>0, \forall L$, $\mathcal{G}^2(L)$ shares the same convexity or monotonicity as $\mathcal{G}(L)$. 
\end{IEEEproof}

With {\em Corollary} \ref{cor1}, the aim of finding the optimum cluster size for the maximized beamforming gain is equivalent to the aim of finding the maximum of $\mathcal{G}(L)$. However, the assumption of having a large $M$ is sometimes not practical. 
This is because the optimum cluster size might be $L^\star=1$ for users located closer to the ELAA, where a practical AP might not have a sufficiently large number of transmit antennas to average out the random term in \eqref{eq10}.
Therefore, a more extensive study of the cluster size or coverage optimization is required. 

\subsection{Optimization for eMBB}
The quality of eMBB service is often related to the average SNR (aSNR) instead of the iSNR. 
Hence, for the eMBB, we are more interested in the average beamforming gain, which can be computed by
\begin{equation}\label{eq16}
\mathbb{E}(\bar{\beta})=\frac{\alpha^2_\bot}{(1+\kappa)}\Big(\frac{\kappa\mathcal{G}^2(L)}{M_\textsc{ap}}+\Gamma(L)\Big),
\end{equation}
where 
\begin{equation}\label{eq17}
\Gamma(L)=\frac{1}{L}\sum_{l=0}^{L-1}(1+l^2\eta)^{-q}.
\end{equation}
Following the proof in \eqref{app06}, it can be easily observed that $\Gamma(L)$ is a monotonically decreasing function. 
Hence, $\mathbb{E}(\bar{\beta})$ can be either a concave function or a monotonically decreasing function, depending on the convexity of $\mathcal{G}(L)$. 

Based on the function $\mathbb{E}(\bar{\beta})$, it is very difficult to find the optimum result ($L^\star$) in a closed form. Nevertheless, we can easily find the optimum cluster size through line search (starting from $L=1$). 

\subsection{Optimization for URLLC}\label{secIIIc}
URLLC is fundamentally different from eMBB in the sense that its performance is dependent on the iSNR or equivalently the instantaneous beamforming gain $\bar{\beta}$ in this study\cite{wang2021massivemimo}. It is shown in \eqref{eq10} that $\bar{\beta}$ is a random variable, and its variation will result in the outage probability formulated below
\begin{equation}\label{eq18}
\mathscr{P}_\mathrm{out}=\int_0^{\bar{\beta}_{\top}}p_{\bar{\beta}}(x)dx,
\end{equation}
where $\bar{\beta}_{\top}$ is the threshold of the instantaneous beamforming gain, and $p_{\bar{\beta}}(x)$ is the probability density function of $\bar{\beta}$. 
URLLC has a stringent requirement for the outage probability (e.g., $10^{-5}$ or even lower), and such imposes an upper bound on $\bar{\beta}_{\top}$.
For the notation simplicity, $\bar{\beta}_{\top}$ is reused for this upper bound. 
Then, for the URLLC, the aim of cluster size optimization is no longer to maximize $\bar{\beta}$ or $\mathbb{E}(\bar{\beta})$. Instead, it is much more important to maximize the upper bound $\bar{\beta}_{\top}$ for the sake of guaranteed reliability and transmit energy efficiency.

Our analysis starts from a compact form of \eqref{eq10}
\begin{equation}\label{eq19}
\bar{\beta}=\frac{\kappa\alpha^2_\bot}{(1+\kappa)M_\textsc{ap}}\mathcal{G}^2(L)+\omega,
\end{equation}
where $\omega$ introduces the randomness of $\bar{\beta}$.
Recall the expression \eqref{eq06}, $\omega$ can be expressed by
\begin{IEEEeqnarray}{ll}\label{eq20}
\omega&=(ab^*+a^*b+bb^*)/M,\\
&\approx(ab^*+a^*b)/M,\label{eq21}
\end{IEEEeqnarray}
where the randomness comes from $b$, and the approximation in \eqref{eq20} is due to the drop of higher-order component of the randomness as it does not dominate the performance. Then, $\omega$ is approximately a white Gaussian process with zero mean and the variance
\begin{equation}\label{eq22}
\sigma^2_\omega=\frac{4a^2\mathbb{E}(\Re^2(b))}{L^2M_\textsc{ap}^2}.
\end{equation}
Applying the above conclusion into \eqref{eq18}, we can obtain the following approximation \cite{Digital_Comm}
\begin{IEEEeqnarray}{ll}\label{eq23}
\mathscr{P}_\mathrm{out}&\approx\frac{\bar{\beta}_{\top}p_{\bar{\beta}}(x=\bar{\beta}_{\top})}{2},\\
&\approx\frac{\bar{\beta}_{\top}}{2\sqrt{2\pi}\sigma_\omega}\exp\Big(
-\frac{(\bar{\beta}_\top-\mathbb{E}(\bar{\beta}))^2}{2\sigma_\omega^2}\Big),\label{eq24}
\end{IEEEeqnarray}
where the approximation is sufficiently accurate for a very small $\mathscr{P}_\mathrm{out}$ such as the one specified for the URLLC.
Given a $\mathscr{P}_\mathrm{out}$, $\bar{\beta}_{\top}$ is a function of $\mathbb{E}(\bar{\beta})$ and $\sigma_\omega^2$.

\begin{figure}[t]
\centering
\includegraphics[scale=0.26]{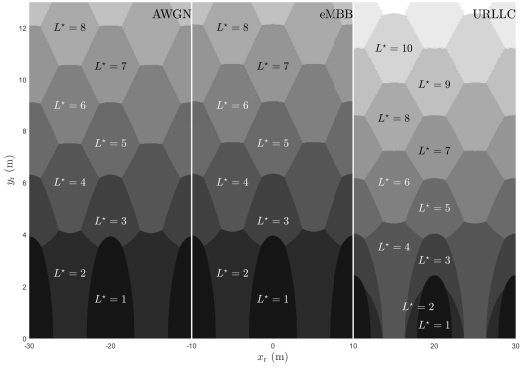}
\caption{Beamforming gain and optimized cluster size of eMBB and URLLC for the Rician K-factor: $\kappa=5$.}
\vspace{-1em}
\label{fig3}
\end{figure}

Mathematically, we cannot have a closed-form solution to \eqref{eq24}. 
Nevertheless, $\bar{\beta}_{\top}$ can be computed through the brute force algorithm.
Moreover, it is worth noting that, for users located far from the ELAA, a larger $L$ will be required to maximize $\mathbb{E}(\bar{\beta})$. At the meantime, \eqref{eq22} implies that the variance $\sigma^2_\omega$ also decreases. It implies that URLLC should have a similar behavior with the eMBB in terms of the cluster size optimization. 
Therefore, line search can also be employed to find $L^\star$ which maximizes $\bar{\beta}_{\top}$.

\section{Simulation Results and Coverage Analysis}\label{secIV}
Computer simulations have been carried out to elaborate our theoretical work presented in Section \ref{secIII}. The system model for computer simulations is the one illustrated in Fig. \ref{fig1}. Basically, there are a set of APs ($L$) with each having a ULA with the size: $M_\textsc{ap}=8$ or $16$.  
The distance between any two neighboring APs is: $d_\textsc{ap}=10$ m. APs operate at the carrier frequency of $3.5$ GHz as specified in the European standards for eMBB and URLLC \cite{Euro5GObservatory}, and they collaboratively provide data services to a user located at the 2D location with the coordinate $(x_\mathrm{r}, y_\mathrm{r})$. The Rician K-factor of the ELAA channel is set to: $\kappa=1, 5,$ or $\infty$ (AWGN), respectively. The path-loss exponent is set to: $q=1$ (free-space).
For the URLLC, the outage probability is specified by: $\mathscr{P}_\mathrm{out}=0.5\times10^{-6}$.
For the purpose of numerical evaluation, we illustrate the optimum cluster size (i.e., $L^\star$) on the 2D map. 
Moreover, the service coverage is evaluated using the coverage probability, which is defined as the probability to achieve the required beamforming gain when a user moves along the direction in parallel to the network-ELAA \cite{8761131}. Thanks to the periodic property of the network-ELAA, the coverage probability $\mathscr{P}_\mathrm{cov}$ can be mathematically described by
\begin{equation}\label{eq25}
\mathscr{P}_\mathrm{cov}\triangleq\int_{X_o}^{X_o+d_\textsc{ap}}\mathscr{P}\left(\bar{\beta}>\bar{\beta}_\mathrm{req}|x_\mathrm{r},y_\mathrm{r}\right)\mathrm{d}x_\mathrm{r},
\end{equation}
where $X_o$ stands for an arbitrary starting point, and $\bar{\beta}_\mathrm{req}$ for the required beamforming gain.

\subsection{Optimized Cluster Size}
\begin{figure}[t]
	\centering
	\includegraphics[scale=0.26]{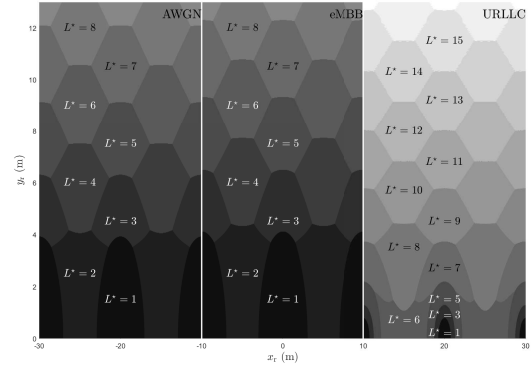}
	\caption{Beamforming gain and optimized cluster size of eMBB and URLLC for the Rician K-factor: $\kappa=1$.}
	\vspace{-1em}
	\label{fig4}
\end{figure}

Fig. \ref{fig3} demonstrates the optimum cluster size as well as the corresponding beamforming gain for the Rician K-factor $\kappa=5$ or $\infty$ (AWGN), respectively. The beamforming gain shows periodic behavior along the x-axis, and thus we only plot $2$ periods for each case of demonstration, i.e., eMBB, URLLC and AWGN \footnote{eMBB and URLLC share the same result for the AWGN case.}.  
Overall speaking, the optimum cluster size ($L^\star$) increases when the user's location moves away from the ELAA. 
This phenomenon confirms our theoretical prediction presented in Section III, i.e., $L^\star$ would grow with the decrease of $\eta=(d_\textsc{ap}/d_\bot)^2$. When the distance between the user and the ELAA becomes sufficiently small ($y_\mathrm{r}<4$ m), the optimum cluster size reduces to $L^\star=1$ (or $L^\star=2$ for places closer to the middle of two neighboring APs). 
This result is also well in line with our theoretical prediction in {\em Theorem} \ref{thm02}. 
To be more specific, we let $d_\bot=y_\mathrm{r}<4$ m and obtain $\eta>6.25$. This result does not satisfy the inequality \eqref{eq14}, and in this case the beamforming gain $\bar{\beta}$ becomes a monotonically decreasing function of $L$. Hence, the maximum beamforming gain is achieved at $L^\star=1$. 
It is not surprising to see $L^\star=2$ or more generally an even number for places closer to the middle of two neighboring APs. This is mainly because APs should provide symmetric contribution to the beamforming gain (rigorous proof omitted for the sake of spacing).  

Fig. \ref{fig3} also shows that, for the case of $\kappa=5$, URLLC generally requires a larger cluster size than eMBB in order to achieve the required beamforming gain. This observation is fully in line with the theoretical analysis in Section \ref{secIIIc}, i.e., more APs are needed in URLLC to reduce the variance $\sigma_\omega^2$. For the case of $\kappa\to\infty$, we naturally have $\sigma_\omega^2=0$, for which URLLC and eMBB share the same result. 
\begin{figure}[t]
	\centering
	\includegraphics[scale=0.27]{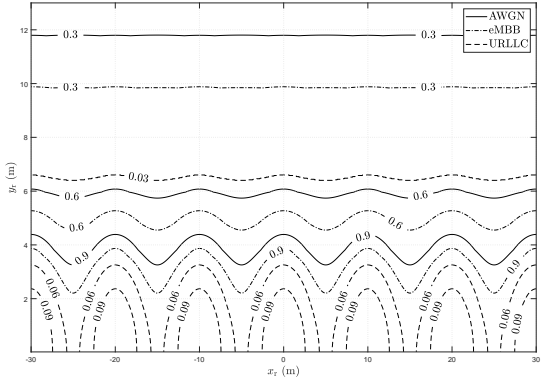}
	\caption{The optimized beamforming gain in a contour form for $\kappa=5$ and $M_\textsc{ap}=16$.}
	\vspace{-0.7em}
	\label{fig5}
\end{figure}
\begin{figure}[t]
	\centering
	\includegraphics[scale=0.27]{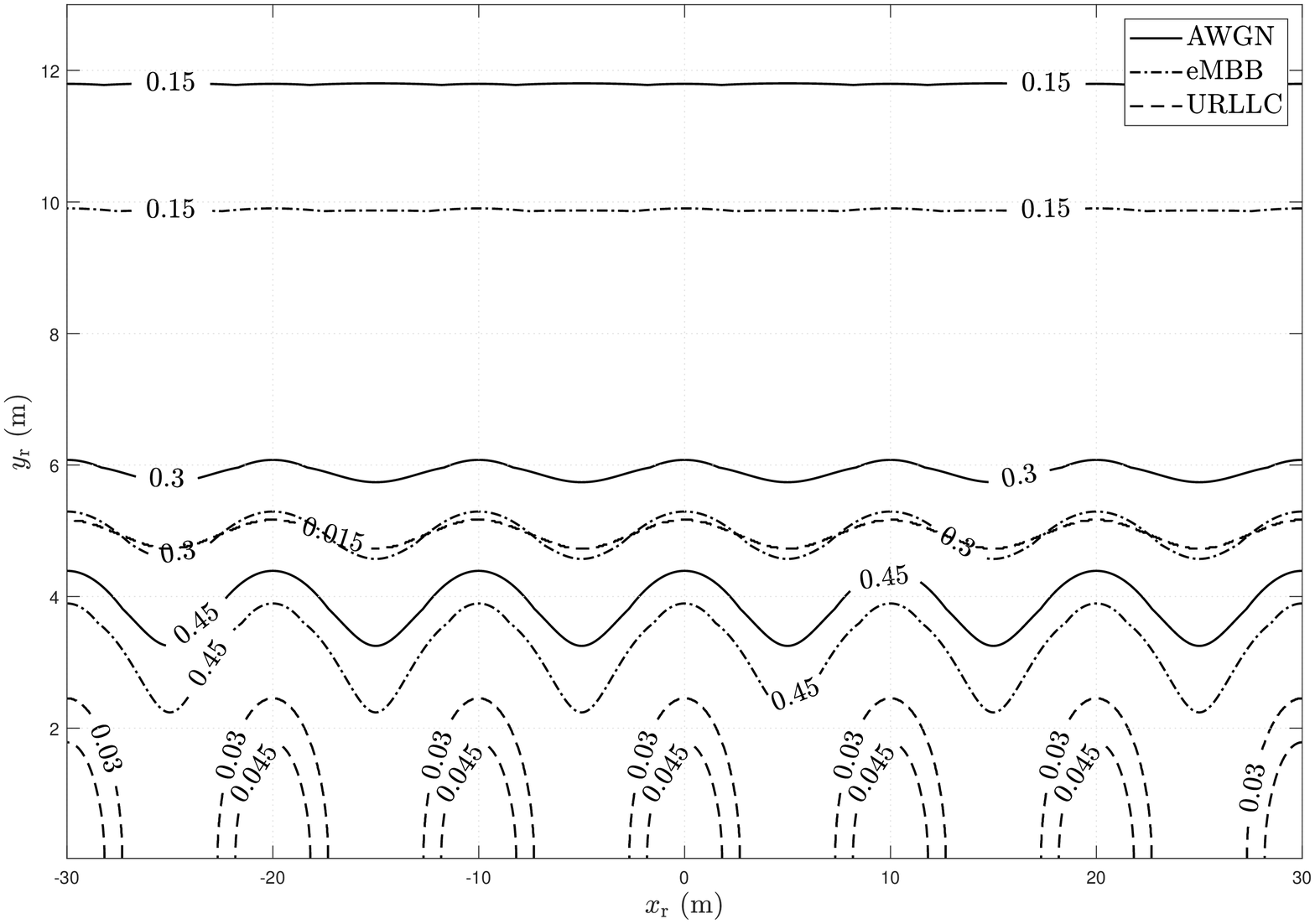}
	\caption{The optimized beamforming gain in a contour form for $\kappa=5$ and $M_\textsc{ap}=8$.}
	\vspace{-0em}
	\label{fig6}
\end{figure}

\begin{figure}[t]
	\centering
	\includegraphics[scale=0.27]{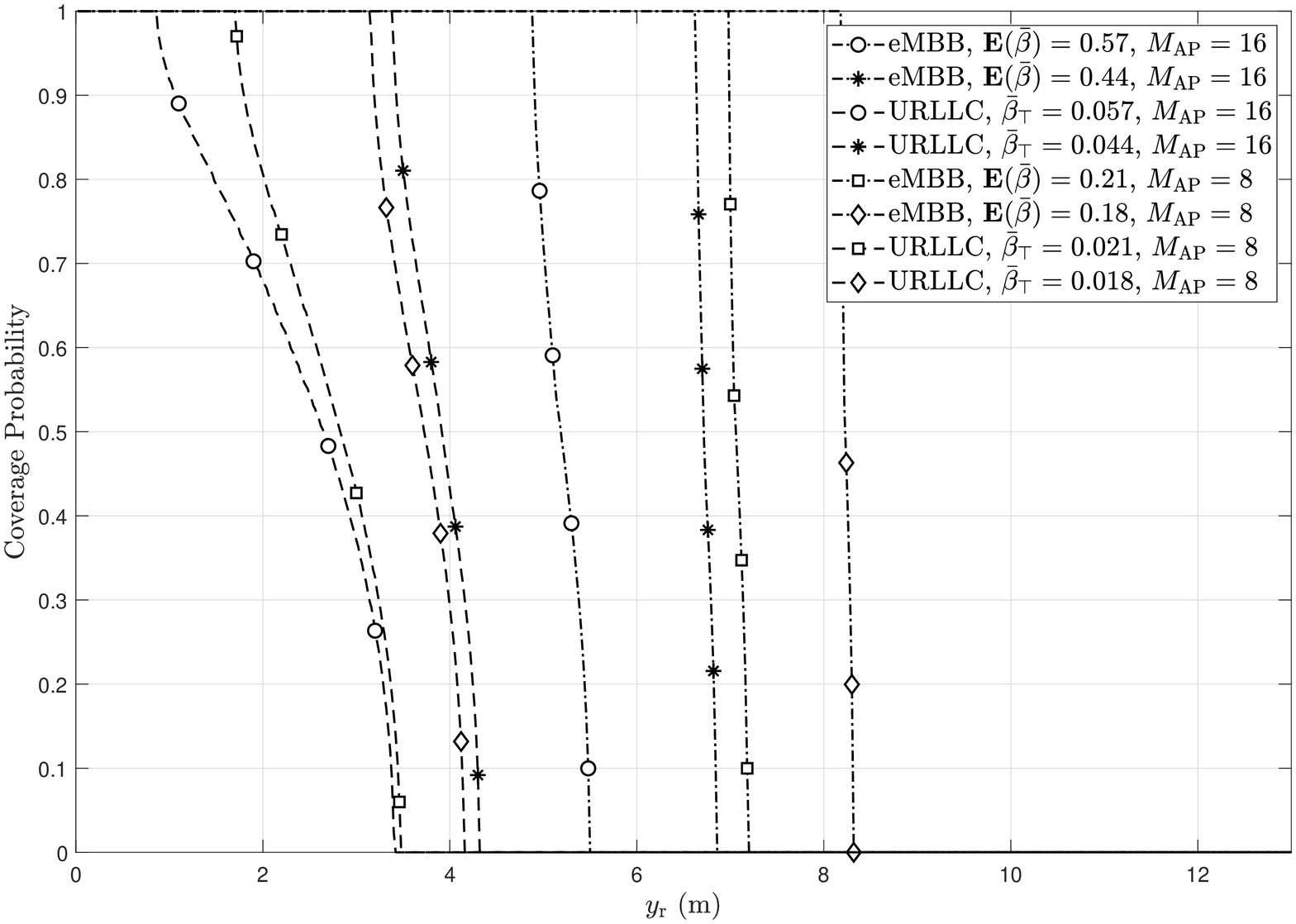}
	\caption{Coverage probability in the case of optimized cluster size when $\kappa=5$, $M_\textsc{ap}$ is $16$ or $8$.}
	\vspace{-0.7em}
	\label{fig7}
\end{figure}

When the Rician K-factor becomes smaller ($\kappa=1$), Fig. \ref{fig4} shows that it does not bring considerable impact to the eMBB. 
This is because the optimization for eMBB is based on the average beamforming gain (i.e., \eqref{eq16}), which does not change considerably with the Rician K-factor.
However, this is not the case for URLLC. As shown in \eqref{eq08} and \eqref{eq22}, the decrease of Rician K-factor results in the increase of the T-CSI uncertainty (measured by the variance $\sigma^2_\omega$), which becomes the dominating factor of the beamforming gain. Then, URLLC will need more APs (effectively more transmit antennas) to reduce the variance $\sigma^2_\omega$. This conclusion is well reflected from Fig. \ref{fig4}.
\begin{figure}[t]
	\centering
	\includegraphics[scale=0.27]{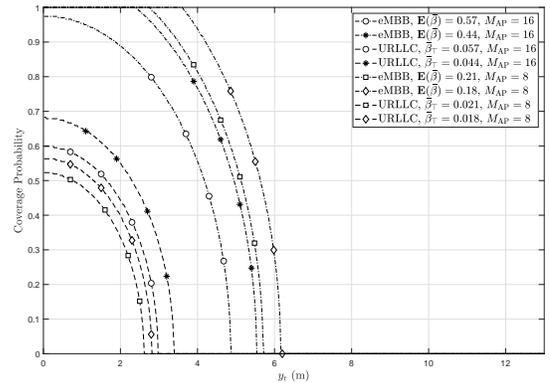}
	\caption{ Coverage probability in the case of single-AP when $\kappa=5$, $M_\textsc{ap}$ is $16$ or $8$.}
	\vspace{-0em}
	\label{fig8}
\end{figure}
\subsection{Coverage Analysis for eMBB and URLLC}

To facilitate the analysis of radio coverage, we plot in Fig. \ref{fig5} and Fig. \ref{fig6} the optimized beamforming gain in a contour form. 
For the sake of spacing, we only show the result for $\kappa=5, \infty$. First of all, both eMBB and URLLC show their maximum coverage in AWGN ($\kappa\to\infty$), as this is the case when the network-ELAA has the ideal T-CSI. When the Rician K-factor reduces to $\kappa=5$, eMBB shows some coverage degradation (around $20\%$) due to the presence of T-CSI uncertainty. Nevertheless, the ELAA can provide almost seamless eMBB coverage along the direction of x-axis. It can also be observed that the URLLC coverage gets significantly degraded due to the presence of T-CSI uncertainty. More severely, the ELAA can no longer offer seamless coverage if solely from the beamforming gain point of view. Certainly, we can improve the coverage by means of power boosting. However, we argue that this is an energy unfriendly approach and could potentially bring more power amplifier nonlinearities into the ELAA system. When comparing the results between Fig. \ref{fig5} and Fig. \ref{fig6}, it is perhaps better to increase the number of transmit antennas per AP for the coverage extension (around $3$ dB). 

In Fig. \ref{fig7} (optimized cluster size) and Fig. \ref{fig8} (single AP), we plot the coverage probability for URLLC and eMBB using the formula \eqref{eq25}. Generally, for the case of optimized cluster size, the eMBB coverage is less sensitive to the distance between the user and the ELAA (i.e., $y_\mathrm{r}$), while the URLLC coverage does. When comparing the ELAA coverage with the single-AP coverage in Fig. \ref{fig8}, we can observe a significant gain of coverage extension  ($50\%$ or more) for most of URLLC and eMBB cases. This conveys a rather encouraging message to both the academic and industrial communities in this domain.

\section{Conclusion and Outlook}\label{sec05}
In this paper, an intensive study on the network-ELAA (or network non-ULA) beamforming and radio coverage for eMBB and URLLC services have been presented. It has been shown that, due to the wireless channel spatial non-stationarity, network-ELAA behaves very differently from the conventional network-MIMO systems in terms of the AP cluster optimization. There exists an optimum cluster size which maximizes the beamforming gain. Moreover, URLLC also behaves differently from eMBB in the sense that it requires more APs to reduce the impact of T-CSI uncertainty inherent in the location-aware ELAA beamforming. Finally, it has been shown that the network-ELAA can offer significant coverage extension ($50\%$ or more in most of cases) when comparing to the single-AP scenario. This is a rather appealing result which encourages further research towards this direction by considering more complicated scenarios involving multiuser, mobility, and radio resource allocation problems.

\appendices
\section{Proof of Theorem \ref{thm01}}\label{appdx1}
Given the condition c1), we can have $P_o=(P)/(M)$ and rewrite \eqref{eq03} into
\begin{equation}\label{app01}
\mathrm{isnr}=\left(\frac{\beta}{M}\right)\left(\frac{P}{\sigma^2}\right).
\end{equation}
Given the condition c2), \eqref{eq06} becomes
\begin{equation}\label{app02}
\beta=\frac{M^2\alpha_0^2}{1+\kappa}\Big|\sqrt{\kappa}+\frac{1}{M}\sum_{m=0}^{M-1}\nu_m\phi_m^*\Big|^2.
\end{equation}
Plugging \eqref{app02} into \eqref{app01} yields 
\begin{equation}\label{app03}
\mathrm{isnr}=\frac{M\alpha_0^2}{1+\kappa}\Big|\sqrt{\kappa}+\frac{1}{M}\sum_{m=0}^{M-1}\nu_m\phi_m^*\Big|^2\left(\frac{P}{\sigma^2}\right).
\end{equation}
For $M\rightarrow\infty$, we have
\begin{equation}\label{app04}
\lim_{M\rightarrow\infty}\frac{1}{M}\sum_{m=0}^{M-1}\nu_m\phi_m^*=\mathbb{E}(\nu\phi^*)=\mathbb{E}(\nu)\mathbb{E}(\phi^*)=0{\color{blue},}
\end{equation}
\begin{equation}\label{app05}
\lim_{M\rightarrow\infty}M\alpha^2_0=1~(\text{free~space}; \text{see} \text{\cite{Rappaport96}}).
\end{equation}
Applying \eqref{app04}-\eqref{app05} into \eqref{app03} results in \eqref{eq09}.

\section{Proof of Theorem \ref{thm02}}\label{appdx2}
We start from the function $f(l)\triangleq(1+l^2\eta)^{-\frac{q}{2}}$, whose partial derivative with respect to $l$ is given by
\begin{equation}\label{app06}
	\frac{\partial f(l)}{\partial l}=-\frac{ql}{(1+l^2\eta)^{(1+\frac{q}{2})}}<0.
\end{equation}
Hence, $f(l)$ is a monotonically decreasing function of $l$.
It means that the following average is a monotonically decreasing function of $L$
\begin{equation}\label{app07}
	\bar{f}(L)=\frac{1}{L}\sum_{l=0}^{L-1}f(l).
\end{equation}
Moreover, it is easy to justify $\bar{f}(L)<(1)/(L), \forall L>1$. 
This means that $\bar{f}(L)$ decreases more than linearly with $L$. 
Then, the function $\mathcal{G}(L)=\sqrt{L}\bar{f}(L)$ would be a decreasing function for a sufficiently large $L$. 
The only possibility to have $\mathcal{G}(L)$ increasing with $L$ is at the beginning stage of $L$. 
Indeed, it can be found that $\mathcal{G}(L)$ is approximately an inverse Gaussian distribution function (rigorous proof omitted), 
which is either a concave function or monotonically decreasing function. For this reason, when the condition $\mathcal{G}(2)<\mathcal{G}(1)$ is satisfied, $\mathcal{G}(L)$ will be a monotonically decreasing function. For the case of $\mathcal{G}(2)>\mathcal{G}(1)$, $\mathcal{G}(L)$ grows at the beginning and then decreases, forming a concave function. Solving the inequality $\mathcal{G}(2)>\mathcal{G}(1)$ leads to \eqref{eq14}.

\section*{Acknowledgement}
This work is partially funded by the 5G Innovation Centre and the 6G Innovation Centre.

\ifCLASSOPTIONcaptionsoff
\newpage
\fi

\bibliographystyle{IEEEtran}
\bibliography{Bib_URLLC,Bib_Else}		
\end{document}